# Ising superconductivity in noncentrosymmetric bulk NbSe$_2$

Dominik Volavka[1], Zuzana Pribulová[2], Jozef Kačmarčík[2], Timon Moško[1], Branislav Stropkai[1], Jozef Bednarčík[1,2], Yingzheng Gao[3], Owen Moulding[3], Marie-Aude Méasson[3], Christophe Marcenat[4], Thierry Klein[3], Shunsuke Sasaki[5], Laurent Cario[5], Martin Gmitra[1,2], Peter Samuely[2], Tomas Samuely[1,✉]

[1]Centre of Low Temperature Physics, Faculty of Science, Pavol Jozef Šafárik University in Košice, 04001 Košice, Slovakia

[2]Centre of Low Temperature Physics, Institute of Experimental Physics, Slovak Academy of Sciences, 04001 Košice, Slovakia

[3]Univ. Grenoble Alpes, CNRS, Grenoble INP*, Institut Néel, 38000 Grenoble, France

[4]Univ. Grenoble Alpes, CEA, Grenoble INP, IRIG, Pheliqs, 38000 Grenoble, France

[5]Institut des Matériaux Jean Rouxel, Université de Nantes and CNRS-UMR 6502, Nantes 44322, France

✉e-mail: tomas.samuely@upjs.sk

## Abstract:

Ising superconductivity allows in-plane upper critical magnetic fields to vastly surpass Pauli limit by locking the antiparallel electron spins of Cooper pairs in the out-of-plane direction. It was first explicitly demonstrated in fully two-dimensional monolayers of transition metal dichalcogenides with large spin-orbit coupling and broken inversion symmetry. Since then, several studies have shown that it can be present in layered bulk materials, too. In our previous study, we have clarified the underlying microscopic mechanism of Ising superconductivity in bulk, based on a reduced electronic coupling between superconducting layers due to intercalation by insulating layers and restricted inversion symmetry. But earlier studies suggest that in some transition metal dichalcogenide polytypes Pauli paramagnetic limit is violated even without intercalation. Here, using heat capacity measurements we unambiguously demonstrate, that the pristine noncentrosymmetric bulk 4H$_a$-NbSe$_2$ polytype significantly violates the Pauli limit. The band structure parameters obtained from ab initio calculations using the experimentally determined crystal structure are used in the theoretical model which provides the microscopic mechanism of the Ising protection based solely on broken inversion symmetry.

Two-dimensional (2D) systems allow the realization of unique quantum phenomena unattainable in the common three-dimensional (3D) world. E. g., in conventional bulk superconductors, an external magnetic field can break the Cooper pairs via the orbital pair breaking. This occurs when magnetic field acts on the Cooper pair's electrons' charge. The Cooper pair breaks when the induced Lorentz force exceeds its binding force. However, by reducing the dimensionality, a 2D superconductor in an in-plane magnetic field is impervious to this so-called orbital pair breaking.

Here, the only remaining effect is the spin pair breaking; the Cooper pair is destroyed by aligning both its antiparallel spins with the magnetic field via the Zeeman effect. The minimum field necessary for spin pair breaking is known as the Pauli paramagnetic limit $B_P$. For materials with negligible spin-orbit scattering, it is[1–3]:

$$B_P = \frac{\Delta}{\sqrt{2}\mu_B} = \frac{1.764 k_B T_c}{\sqrt{2}\mu_B} \approx 1.86 \times T_c [T/K] \tag{1}$$

where $\Delta$ is the superconducting energy gap, $\mu_B$ is the Bohr magneton, $k_B$ is the Boltzmann constant, and $T_c$ is the superconducting critical temperature. Moreover, in a 2D superconductor with an in-plane mirror symmetry, the spins of Cooper pairs are locked perpendicular to the plane. Further, if spin-orbit coupling is strong, breaking the inversion symmetry lifts the Kramers degeneracy. Consequently, the electrons with opposite momenta constituting a Cooper pair are subject to opposite effective intrinsic Zeeman fields. This prevents spin pair breaking even at external magnetic fields larger than $B_P$[4–6]. This phenomenon – dubbed Ising superconductivity (IS) – was experimentally demonstrated and theoretically explained in non-centrosymmetric 2D superconductors [7–9]. In the epitomical monolayer NbSe$_2$, the in-plane magnetic field $B_{c2}^{\parallel}$ overcomes $B_P$ seven times.

Perhaps more exciting than the enhanced resilience against magnetic fields is the potential application of IS in realizing various exotic phenomena such as equal spin Andreev reflections[10], topological superconductivity[11–14], and Majorana fermions[10,15,16].

Albeit a fascinating phenomenon, the assumed restriction to 2D structures would render it challenging for practical applications as well as in depth analysis compared to regular 3D materials. Bulk materials are generally more robust, easily scalable and accessible to a larger range of scientific analytical techniques. However, by stacking layers of NbSe$_2$, the IS is gradually suppressed and already in a few layers disappears completely (mostly due to regained Kramers degeneracy)[7].

Though pristine bulk 2H$_a$-NbSe$_2$ does not exhibit IS, the orbital pair-breaking remains ineffective when the layers are parallel to the applied magnetic field. In this configuration, the superconductivity endures fields up to around $B_P$. The layered structure of the bulk material, with adjacent $ab$ planes bound by van der Waals interactions, protects the 2D effect in bulk [17–22]. On the contrary, in orthogonal geometry the superconductivity ceases well below $B_P$.

A natural question arises: Can IS be protected in bulk? Several recent and older studies confirm this. A straightforward way is to electronically decouple superconducting transition metal dichalcogenide (TMD) layers by intercalation, thus breaking the inversion symmetry and effectively eliminating Kramers degeneracy[23–36]. The microscopic mechanism behind this kind of IS protection was elucidated in our previous study[37]. There, the misfit single crystals (LaSe)$_{1.14}$(NbSe$_2$) with the superconducting transition temperature of $T_c$ = 1.3 K, and (LaSe)$_{1.14}$(NbSe$_2$)$_2$ with $T_c$ = 5.3 K comprising intercalated monolayers and bilayers of NbSe$_2$ were investigated. A concerted effect of charge-transfer from LaSe to NbSe$_2$, defects, reduction of interlayer hopping between superconducting NbSe$_2$ layers, and stacking enables Ising superconductivity in these bulk compounds far above $B_P$.

But previous studies[38–42] indicate, that the noncentrosymmetric 4H$_a$ TMD polytypes[43], comprised of juxtaposed TMD monolayers without any intercalation, can violate the Pauli limit, too.

Nonetheless, these studies fail to fully exploit the 3D nature of the materials and utilize experimental techniques that do not offer conclusive evidence of the 3D IS. In fact, purely 2D spurious effects such as surface superconductivity and percolation transport can mimic the real superconductivity in lamellar superconductors. Separated 2D layers with formation of Josephson vortices in magnetic field parallel to the layers could affect the determination of the upper critical magnetic field via transport measurements. The estimates of the extreme upper critical fields in these $4H_a$ polytypes are all based on transport measurements. As shown in previous studies, the $(LaSe)_{1.14}(NbSe_2)$[44] but also $Ba_{0.75}ClTaSe_2$[31], both systems with extreme $B_{c2}^{\parallel}$, can be regarded as stacks of 2D superconducting planes with interlayer Josephson coupling where Josephson vortices penetrate in between the planes. Then, very broadened resistive transitions from zero to the normal state occur and the determination of $B_{c2}^{\parallel}$ is not straightforward. To conclusively prove the Pauli limit violation in non-intercalated bulk TMDs, we assess the upper critical magnetic fields of $4H_a$-$NbSe_2$ by heat capacity measurements and show that $B_{c2}^{\parallel}$ achieves values almost 3 times $B_P$. Then, by analysing the band structure, we validate the relevance of IS protection in the compound.

## Crystal structure

The investigated bulk $4H_a$-$NbSe_2$ sample has the same stoichiometry as the more common $2H_a$-$NbSe_2$ polytype and the difference between them is subtle. Therefore, a meticulous crystallographic characterization of the studied sample is essential. As shown in Fig. 1a, in the $4H_a$ polytype, layers labelled A and B stack in the $2H_a$ form and layer C is translated by a vector $(2a/3+b/3)$ within the $ab$ plane relative to the B layer. This leads to a reduced symmetry from $P6_3/mmc$ ($2H_a$) to $P\bar{6}m2$ ($4H_a$), respectively with and without central inversion symmetry[43], and introduces an out-of-plane mirror symmetry with the mirror $ab$ plane in the middle of layers A and C.

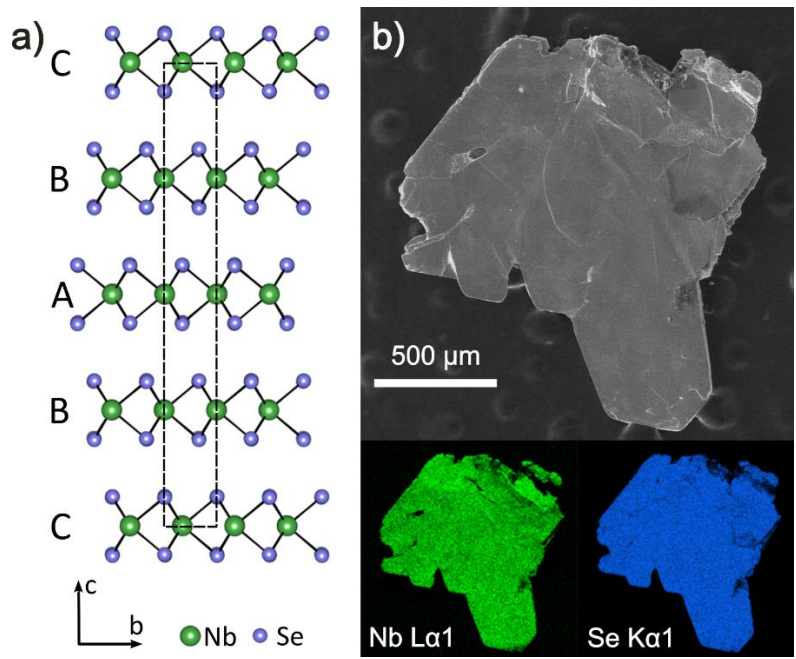

**Figure 1 – 4H$_a$-NbSe$_2$ crystal.** a) Crystal structure of 4H$_a$-NbSe$_2$. Dashed rectangle represents the unit cell in the *bc* plane. b) Backscattered electron image of the pristine cleaved surface of 4H$_a$-NbSe$_2$ crystal (top), elemental mappings of 4H$_a$-NbSe$_2$ surface for Nb and Se (bottom).

## Energy-Dispersive Spectroscopy

We analysed the composition of the sample by a scanning electron microscope (SEM) equipped with an energy-dispersive X-ray spectrometer (EDS) and a backscattered electron (BSE) detector (Fig. 1b). The sample surface was cleaved between several subsequent measurements and no variations in composition were detected. The BSE image shows practically no separation of phases, which was confirmed by elemental mapping. The atomic ratio of Nb and Se obtained from the EDS analysis at different spots is 0.97(3):2. No other elements were detected. The maps of chemical elements indicate their homogeneous distribution. Hence, the SEM analysis confirms that our sample is indeed a layered monocrystal of NbSe$_2$.

## X-ray photoemission spectroscopy

The identical chemical composition is further verified by X-ray photoemission spectroscopy (XPS) shown in Fig. 2a. We directly compared the investigated 4H$_a$-NbSe$_2$ sample with a commercial 2H$_a$-NbSe$_2$ reference sample. Both compounds show identical chemical composition. The XPS survey over a broad range of energies reveals XPS and Auger peaks of Nb, Se, Mo, O and C, with the main XPS peaks labelled. The additional peaks for C, O and Mo correspond to the carbon tape used to attach the sample and the molybdenum sample holder. The comparison of Nb 3d and Se 3d peaks for the two samples shows minor changes in the electronic environment. A detailed description is available in the supplementary information[45].

## X-ray diffraction

For the X-ray diffraction (XRD) measurements, we used a mechanically milled specimen to increase the signal to noise ratio, which was initially low due to the small size of available samples. The pattern we obtained for our investigated sample coincides well with the Rietveld refinement profile for 4H$_a$-NbSe$_2$ determined by Zhou et al.[41] (Fig. 2b), which is distinct from the common 2H$_a$-NbSe$_2$ polytype[46]. Partial reflections at $2\theta$ = 39° and 53°, accounted for by the background in Fig. 2b, do not coincide with the 4H$_a$-NbSe$_2$ profile. We attribute these peaks to defects that presumably correspond to the intergrowth with the 4H$_{dII}$ polytype[47].

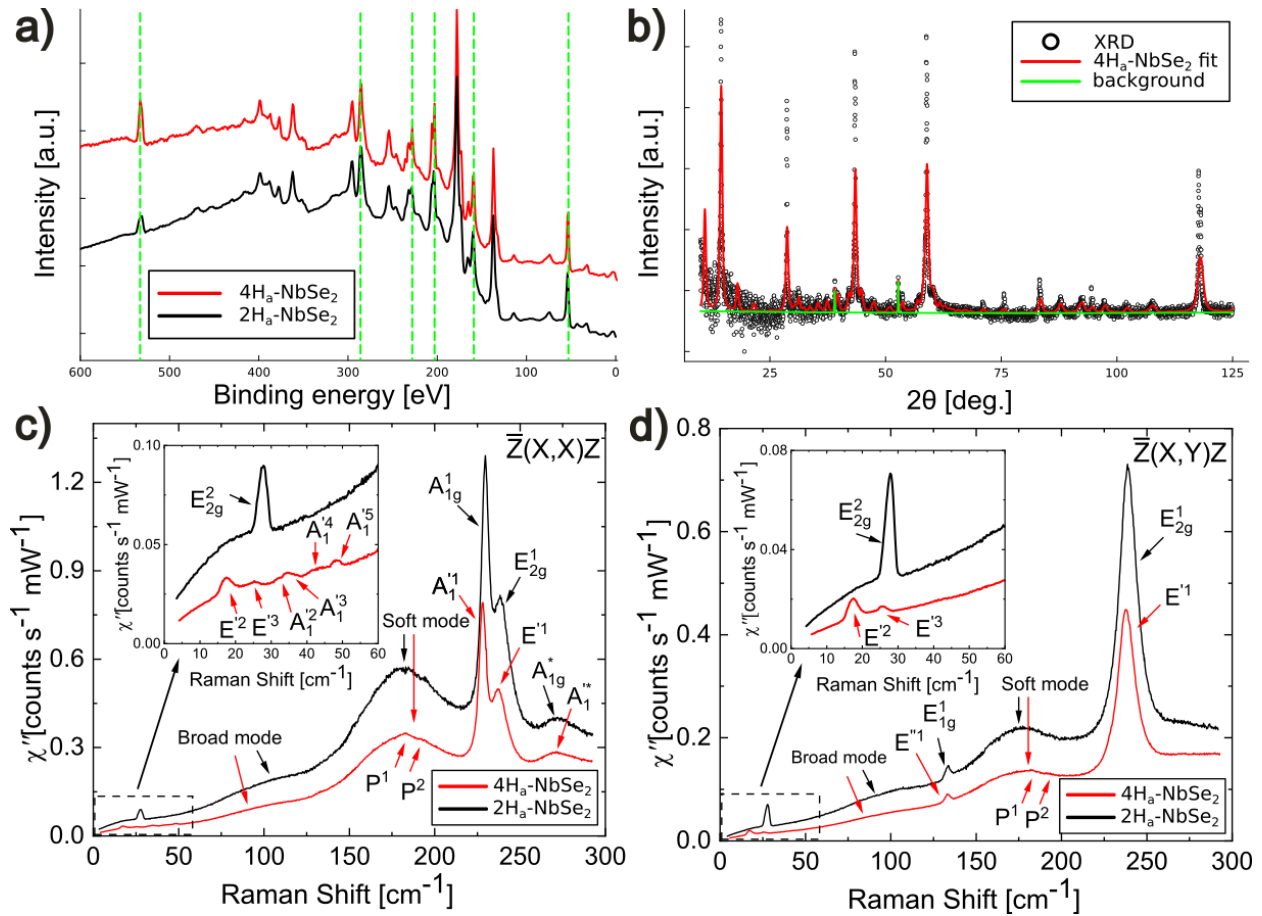

**Figure 2 – X-ray and Raman analysis.** a) XPS of 2H$_a$-NbSe$_2$ and 4H$_a$-NbSe$_2$. The main peaks denoted by the green dashed lines correspond to (from left to right): O1s, C1s, Mo3d, Nb3d, Se3p and Se3d. b) Circles represent the XRD pattern measured on the milled sample. Red line is the Rietveld refinement profile calculated by the GSAS-II software[48] using the crystallographic data for 4H$_a$−NbSe$_2$ determined by Zhou et al[41]. Green line is the background. c) Raman spectra of 2H$_a$-NbSe$_2$ and 4H$_a$-NbSe$_2$ at room temperature in $\overline{Z}(X,X)Z$ configuration, where incident light and scattered light are parallel to one another. d) Raman spectra in $\overline{Z}(X,Y)Z$ configuration, where incident light and scattered light are perpendicular to one another. Black and red arrows mark the intense modes of 2H$_a$-NbSe$_2$ and 4H$_a$-NbSe$_2$, respectively. Insets show the low energy part of spectra marked by the dashed rectangle.

Raman spectroscopy

Fig. 2 c, d present the room temperature Raman spectra of 2H$_a$-NbSe$_2$ and 4H$_a$-NbSe$_2$ in $\overline{Z}(X,X)Z$ and $\overline{Z}(X,Y)Z$ configurations. At high energy (larger than 60 cm$^{-1}$) the Raman responses of 2H$_a$-NbSe$_2$ and 4H$_a$-NbSe$_2$ are generally similar. The most intense modes can be characterized as: two phonons, namely $A_{1g}^1(A_1'^1)$ and $E_{2g}^1(E'^1)$ at ~ 230 cm$^{-1}$, a broad mode at ~ 180 cm$^{-1}$ which is a two-phonon Raman feature (generally identified as a soft mode), $E_{1g}^1(E''^1)$ phonon at 130 cm$^{-1}$, a broad mode at ~ 100 cm$^{-1}$ that appears in both configurations, and an additional broad mode $A_{1g}^*(A_1'^*)$ at ~ 260 cm$^{-1}$ which is probably a double phonon one. For 4H$_a$-NbSe$_2$, two additional phonons ($P^1$ and $P^2$) at ~ 180cm$^{-1}$ are observed in both $\overline{Z}(X,X)Z$ and $\overline{Z}(X,Y)Z$ configurations. This could indicate that these modes belong to the $E'$ symmetry. However, they persist in the pure $A_1'$ symmetry spectra, obtained from substraction. We name them $P$ modes and leave their assignment open. At low energy, 4H$_a$-NbSe$_2$ behaves differently from 2H$_a$-NbSe$_2$. We observe six phonons between 15 cm⁻¹ and 50 cm⁻¹ in the 4H$_a$ structure and the single shear mode in the 2H$_a$ one. According to their symmetries they are assigned as $E'^2$, $E'^3$, $A_1'^2$, $A_1'^3$, $A_1'^4$ and $A_1'^5$. By comparison with the response of the 2H$_a$ compound, we attribute the most intense $E'$ symmetry phonon ($E'^2$ at ~ 20 cm⁻¹) to the inter-layer shear mode of 4H$_a$-NbSe$_2$.

In 2H$_a$-NbSe$_2$ which has a $P6_3/mmc$ space group (#194, $D_{6h}$ point group), there are 4 Raman active phonons[49]:

$$\Gamma_{2H_a-NbSe_2} = 1A_{1g} + 2E_{2g} + 1E_{1g} \tag{2}$$

We have observed all 4 Raman active phonons ($A_{1g}^1$, $E_{1g}^1$, $E_{2g}^1$ and $E_{2g}^2$). For 4H$_a$-NbSe$_2$ with $P\bar{6}m2$ space group (#187, $D_{3h}$ point group), there are 16 Raman active phonons:

$$\Gamma_{4H_a-NbSe_2} = 5A_1' + 6E' + 5E'' \tag{3}$$

We have observed 5 $A_1'$ modes ($A_1'^1$, $A_1'^2$, $A_1'^3$, $A_1'^4$ and $A_1'^5$), 3 $E'$ modes ($E'^1$, $E'^2$ and $E'^3$), 1 $E''$ mode ($E''^1$). Without considering $P^1$ and $P^2$, 3 $E'$ modes and 4 $E''$ modes remain to be assigned.

The extracted parameters of the phonon modes and the details of fitting are available in the supplementary information[45]. By comparing the modes of the same nature for 2H$_a$-NbSe$_2$ and 4H$_a$-NbSe$_2$, we found similar energy and width for $A_{1g}^1(A_1'^1)$, $E_{2g}^1(E'^1)$, $E_{1g}^1(E''^1)$, $A_{1g}^*(A_1'^*)$, soft modes and broad modes. However, the energy difference between inter-layer shear modes $E_{2g}^2$ and $E'^2$ is significant and helps to identify the different structures. $E'^3$ has an energy close to the shear mode $E_{2g}^2$ of 2H$_a$-NbSe$_2$, questioning the presence of 2H$_a$-contamination in the 4H$_a$ sample. Nonetheless, the temperature dependence measurements (unpublished) confirm that $E'^3$ behaves differently from $E_{2g}^2$, ruling out this possibility. We can safely claim that the level of contamination of the 2H$_a$ structure inside the 4H$_a$ sample is below 1%.

### Superconducting properties from heat capacity

The crystallographic analysis unambiguously confirmed that our sample is a 4H$_a$-NbSe$_2$ polytype. To obtain its bulk superconducting properties, we perform heat capacity measurements at different temperatures down to 0.7 K and magnetic fields ranging from 0 T to 36 T with the NbSe$_2$ layers oriented either parallel or perpendicular to the applied magnetic field. When magnetic field parallel to the sample's layers is applied, the behaviour of the investigated 4H$_a$ polytype differs significantly compared to the common 2H$_a$ polytype. Fig. 3 reveals that the in-plane upper critical magnetic field determined by the measurements of the heat capacity substantially exceeds the Pauli limit confirming previous reports[38–41]. Fig. 3a shows the total heat capacity of the sample plus addenda *C/T* for the magnetic field oriented parallel to the *ab* planes, measured using the thermocouple up to 8 T. Each curve represents a measurement in the designated fixed magnetic field while sweeping the temperature. All curves feature a clear jump at the transition to the superconducting state. Increasing magnetic field shifts superconducting anomaly towards lower temperatures while gradually smearing the effect. The magnetic field of 8 T was not enough to suppress superconductivity in the sample in the overall temperature range, therefore additional measurements were performed in a high-field laboratory in Grenoble, France. Fig. 3b summarizes measurements of the sample´s heat capacity in magnetic field 8 – 19 T using the Cernox chip as a calorimeter. With increasing magnetic field, the anomaly is shifted to even lower temperatures, and the smearing becomes substantial. Fig. 3c displays *C/T* measurements of the sample while sweeping the magnetic field at a fixed temperature. The measurements at 2.8 K and 3.2 K were performed in 19 T magnet, while those at 1.8 K and 0.7 K in 36 T magnet. The transition in the field sweeps is rather wide, with the width of several Tesla. The curves measured at 0.7 K and 1.8 K also feature quite a strong change of slope at low magnetic fields at around 1-2 T. This is probably related to the unconventional character of the energy gap of the system which will be addressed in the forthcoming study. Fig. 3d shows the total heat capacity divided by temperature of the sample plus addenda *C/T* for the magnetic field oriented perpendicular to the *ab* planes. In this case, the field of 5 T was enough to suppress superconductivity in the overall temperature range. For a complete picture, we also show field sweeps at various fixed temperatures for the field perpendicular to the sample's *ab* planes in Fig. 3e. The superconducting anomaly progressively evolves with increasing temperature, shifting towards lower magnetic fields and augmenting its height. The change of slope at the low field is reproduced for this orientation of the magnetic field, too. Moreover, at low temperatures (see curve at 0.77 K), we observed a nonlinearity between ca. 2 and 3 T. It is related to the vortex dynamics, and it occurs due to the character of the measurement (the magnetic field is swept continuously). We will address this effect in detail in a future study, as it goes beyond the scope of the present paper. From all heat capacity measurements sweeping both temperature and magnetic field, respectively, we determined the upper critical magnetic field as a mid-point of the anomaly at the transition. Derived points are collected in Fig. 3f as full symbols (see Figure legend and caption for details). From the field sweeps in high magnetic fields (Fig. 3c) we also extracted a position of the maximum of the anomaly, depicted as open symbols in Fig. 3f. In the field sweep measured at 0.7 K (the lowest curve in Fig. 3c), the uncertainty in the mid-point determination is too high, therefore only the position of the maximum is inserted in $B_{c2}(T)$ plot (Fig. 3f).

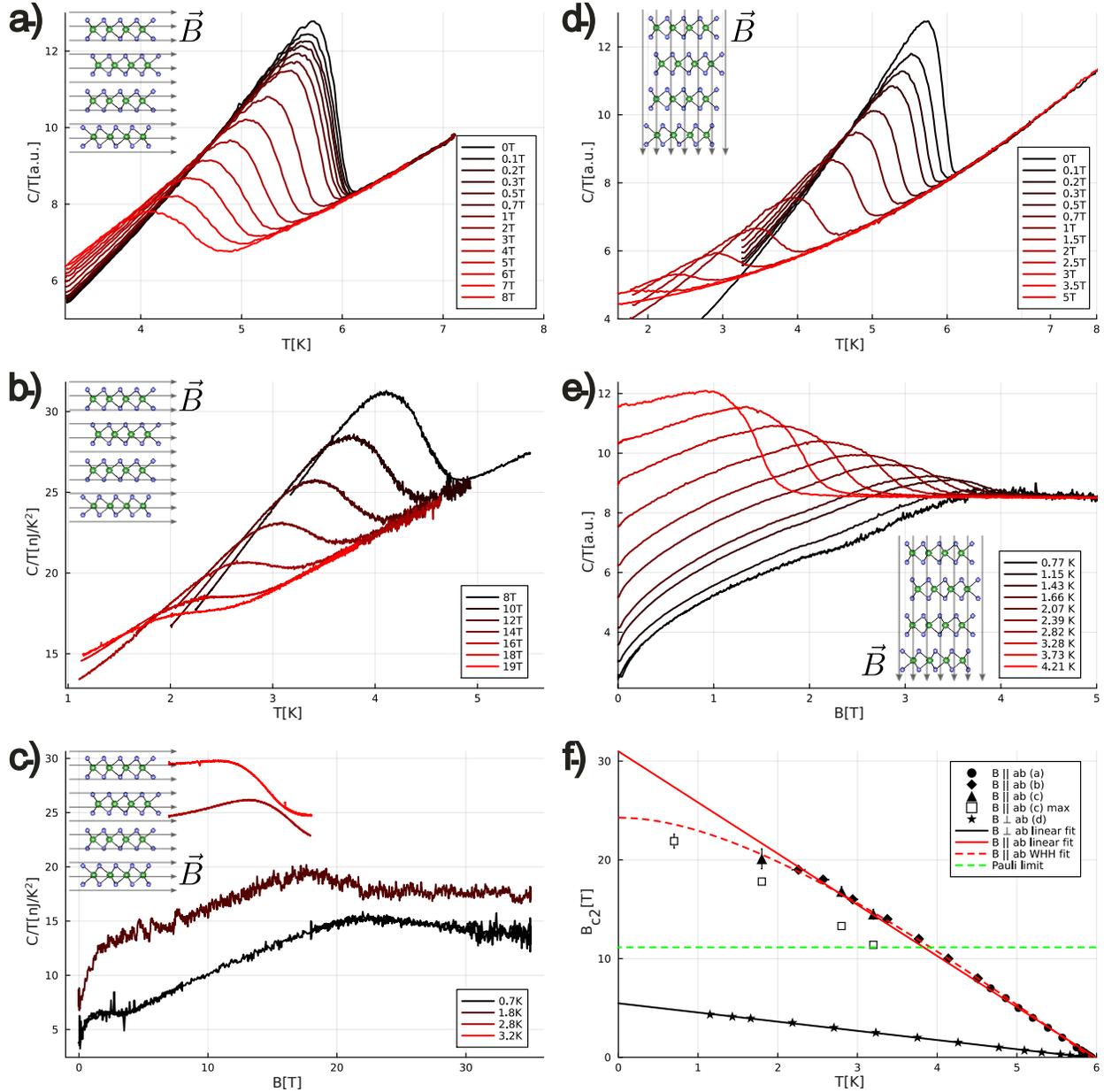

**Figure 3 – Critical magnetic fields of 4H$_a$-NbSe$_2$ determined by heat capacity measurements.** a) Total heat capacity of the sample plus addenda $C/T$ and b) heat capacity of the sample $C/T$ for the magnetic field parallel to the sample's *ab* planes. c) Magnetic field sweeps at various fixed temperatures for the field parallel to the sample's *ab* planes. d) Total heat capacity for the magnetic field oriented perpendicular to the *ab* planes. e) Magnetic field sweeps at various fixed temperatures for the field perpendicular to the sample's *ab* planes. f) Upper critical magnetic fields derived from the heat capacity measurements. Star symbols show $B_{c2}$ for fields applied perpendicular the *ab* planes, other symbols correspond to the parallel alignment. The corresponding linear fits extrapolated to zero temperature are represented by the solid lines. The dashed red line is the theoretical fit using the Werthamer-Helfand-Hohenberg (WHH) formula. Horizontal dashed green line marks the Pauli paramagnetic limit $B_P$. Full symbols were determined at the mid-point of the anomaly at the transition, empty squares at the maximum (not fitted).

From Fig. 3f we hence obtain $T_c \approx 6$ K, which translates to $B_P \approx 11.15$ T. The slightly lower $T_c$ than 7.2 K in 2H$_a$-NbSe$_2$[19,50,51] is in a good agreement with previous studies[39–41,52]. With the layers parallel to the applied magnetic field, we obtain the critical magnetic field $B_{c2}^{\parallel} \approx 31$ T $\approx 2.8 \times B_P$ from the linear fits extrapolated to zero temperature. The extrapolation in line with a standard WHH prediction[53] yields $B_{c2}^{\parallel} \approx 24$ T $\approx 2.2 \times B_P$. In orthogonal geometry, $B_{c2}^{\perp} \approx 5.5$ T, well below $B_P$.

## Band structure with Ising spin-orbit coupling

Hence, from heat capacity experiments we undoubtedly established that the $B_{c2}^{\parallel}$ of bulk 4H$_a$-NbSe$_2$ is well above the Pauli limit. To assess whether IS is a viable explanation, let us first examine its crystal structure. The adjacent NbSe$_2$ layers are directly bound by van der Waals forces as in 2H$_a$-NbSe$_2$, implying comparable interlayer coupling. But the inversion symmetry, which suppresses IS in 2H$_a$-NbSe$_2$, is absent in 4H$_a$-NbSe$_2$. 4H$_a$-NbSe$_2$ can be viewed as a trilayer of NbSe$_2$ in the 2H$_a$ configuration (layers BAB in Fig. 1a), sandwiched between two shifted C layers. It has in-plane mirror symmetry and no inversion symmetry, like the trilayer (and monolayer) NbSe$_2$ where IS is present[7,54]. To further validate the presence of IS, we analyse the band structure of 4H$_a$-NbSe$_2$ obtained from ab initio calculations using the experimentally determined crystal structure (Fig. 4). The bandwidth along $\Gamma - A$ at the Fermi level crossing, which represents the interlayer coupling, is ca. 600 meV, similar to 2H$_a$-NbSe$_2$[55,56]. At the K (and K') point, unlike in 2H$_a$-NbSe$_2$, four pairs of spin-split Nb bands are present. The spin-orbit splitting of the band pairs is ca. 150 meV, like in 2H$_a$-NbSe$_2$. Layer C primarily contributes the topmost pair (light blue). Both B layers equivalently contribute two pairs (orange). Layer A contributes a pair of spin-split bands (red), overlapping the upper B pair. We attribute this overlap to the 2H$_a$ stacking of layers A and B. Notably, since layer A is oriented conversely to layers B and C, the spin polarization of its bands is inverted, as seen in the inset of Fig. 4. Effectively, in addition to the spin-degenerate hybridized AB band, two bands without Kramers degeneracy are present and shifted in energy, enabling IS.

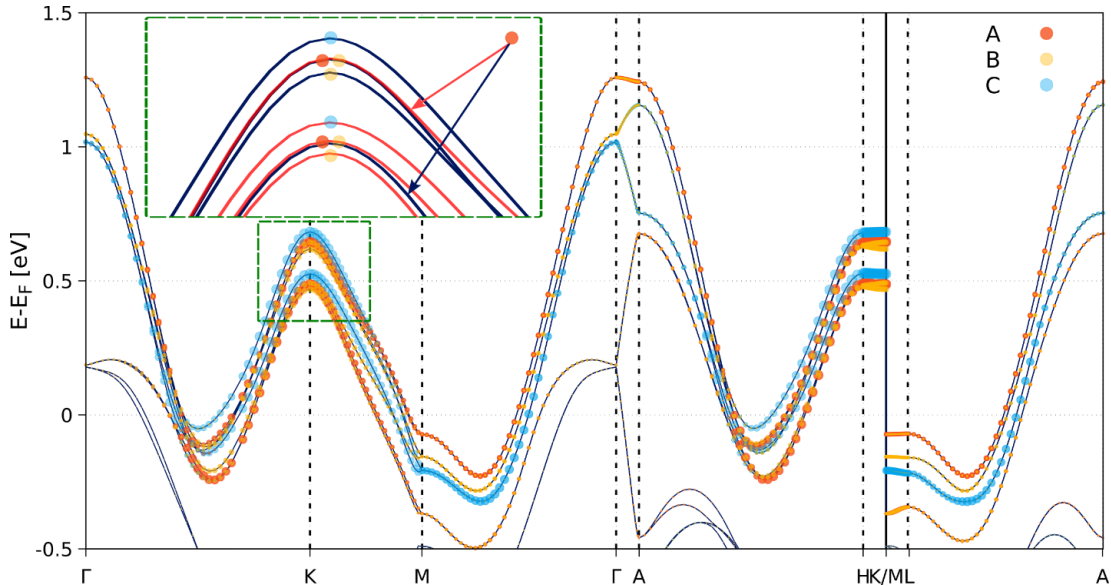

**Figure 4 – Calculated electronic band structure of 4H$_a$-NbSe$_2$.** The circle radii correspond to the Bloch state projection onto the Nb d orbitals of individual layers along high-symmetry lines in the 1$^{st}$ Brillouin zone. Inset: Zoom at the K point indicated by the green dashed rectangle. The colour of the bands indicates the spin polarization projected perpendicular to the layers[45]. Due to symmetry, the in-plane spin polarization is zero.

We can use the band structure parameters to get a rough quantitative estimate of the IS protection. In our previous study[37], we introduced a model where an energy shift of $\pm T_0$ between the NbSe$_2$ spin-split bands lifts Kramers degeneracy. Together with spin-orbit splitting this creates an effective intrinsic magnetic field $B_t$ protecting the Cooper pair spins against external fields parallel to the layers:

$$\mu_B B_t = \sqrt{t_0^2 + (T_0 + \zeta)^2} - \sqrt{t_0^2 + (T_0 - \zeta)^2} \qquad (4)$$

Here, $\zeta$ is half of the spin-orbit splitting, $t_0$ is 1/4 of the bandwidth along the out-of-plane direction and represents the interlayer coupling, $T_0$ is half of the mutual shift of the spin-split bands and $\mu_B$ is the Bohr magneton. The mutual energy shift of the bands is ca. 56 meV at the K (and K'), but it is k dependent. Based on our calculations, we consider 20 meV as the lower bound of the energy shift. Hence, $\zeta \approx 75$ meV, $t_0 \approx 150$ meV, $T_0 \approx 10$ meV. So even though the decoupling of adjacent layers is insignificant, the broken symmetry yields $B_t \approx 154.3$ T, which is almost 14 times larger than the Pauli paramagnetic limit $B_P$. Though likely somewhat reduced by e.g. crystal stacking imperfections that can break the mirror symmetry potentially yielding in-plane (Rashba) spin components[57], this intrinsic effective field provides ample protection of IS.

To summarize, we employed heat capacity measurements to unambiguously demonstrate, that bulk superconductivity in 4H$_a$-NbSe$_2$ single crystal withstands magnetic fields oriented parallel to its layers, that are almost three times larger than $B_P$ – a hallmark of IS. Utilizing crystallographic parameters that we confirmed by multiple experimental techniques, we ab initio calculated the band structure of this distinct NbSe$_2$ polytype. We showed that the broken inversion symmetry of the 4H$_a$-NbSe$_2$ instigates an energy shift between the spin-split pairs of Nb bands of the NbSe$_2$ layers. We employed a previously devised theoretical model and justified IS protection in this noncentrosymmetric bulk NbSe$_2$ polytype without any intercalation. In fact, our calculations indicate, that even an order of magnitude higher critical fields than $B_P$ are possible for such systems. From a more general perspective, our study demonstrates a new way how a peculiar phenomenon associated with 2D systems which are prone to degradation[58] can be carried over to more robust 3D materials. These are not only better for applications but can be scrutinised by a larger number of experimental techniques than their 2D counterparts.

# Methods

## Sample preparation

The single crystal sample of 4H$_a$-NbSe$_2$ was obtained during an attempt to prepare the ternary misfit layered compound (LaSe)$_{1.14}$(NbSe$_2$)$_2$. This synthesis was done in two stages as reported elsewhere[59]. For the first one, a mixture of La/Nb/Se = 1.14/2/5.40 was subject to thermal treatment in a sealed silica tube, initially at 200 °C for 12 h and then 900 °C for 240 h. The second stage consisted in a chemical vapour transport experiment. 500 mg of the obtained black powder was placed in the new silica tube (length: 15 cm) with 43 mg of iodine. The silica tube was then sealed under vacuum and placed in the tubular furnace with the temperature gradient where the reaction mixture was set at 900 °C and the other side was held at 750 °C. The thermal treatment was kept for 240 h, and the reaction was subsequently quenched by plunging the tube into water.

This chemical vapour transport experiment yielded a mixture of (LaSe)$_{1.14}$(NbSe$_2$)$_2$ and NbSe$_2$ crystals. The crystals were sorted out thanks to their different shapes and their composition was subsequently ascertained by energy-dispersive X-ray spectra. Analyses of the NbSe$_2$ crystals did not reveal any trace of La contamination. The reference 2H$_a$-NbSe$_2$ crystal was purchased from HQ Graphene (https://www.hqgraphene.com/).

### Energy-dispersive X-ray spectroscopy (EDS)

The chemical composition of the sample was inspected by energy-dispersive X-ray spectroscopy using a Tescan Vega3 scanning electron microscope operating at 30 keV. The crystal was positioned on a carbon tape.

### X-ray photoemission spectroscopy (XPS)

The calibration of the spectra was based on measuring the binding energies of Au 4f7/2 and Cu 2p3/2 peaks. The XPS spectra of both samples, recorded following the same protocol, were performed by a high-resolution spectrometer Specs PHOIBOS using Al anode with power 250 W. The pressure during measurements was less than 2×10$^{-8}$ mbar. The investigated samples were placed on a molybdenum sample holder using conductive carbon tape and cleaved using scotch tape.

### X-ray diffraction (XRD)

XRD measurements were performed on a mechanically milled crystal in reflection Bragg-Brentano geometry using a Bruker D2 Phaser diffractometer equipped with Cu−Kα radiation. Due to limited amount of the sample material, it was placed onto asymmetrically cut silicon single crystal to minimize the background signal stemming from the sample holder. XRD patterns were recorded by scanning scattering angle 2Θ from 10° up to 125°, with the step size of 0.02°. During such a scan the sample was spinning at the rate of 10 rpm to ensure improved grain statistics. Several independent scans were averaged out to attain good counting statistics. Despite all of this, the peak profiles corresponding to the Bragg peaks of NbSe$_2$ hexagonal phase in measured XRD pattern are suffering from limited grain statistics due to small amount of studied material and preferential orientation of individual powder particles.

### Raman spectroscopy

The polarised Raman spectroscopy experiment is performed at room temperature to study the lattice dynamic of 4H$_a$-NbSe$_2$ and to characterise the quality of the sample. A sample of 2H$_a$-NbSe$_2$ is measured under the same experimental conditions during the same run. We used a 532 nm solid-state laser with an incident laser power of 10 mW and a Trivista spectrometer equipped with ultra-low noise, cryogenically cooled PyLon CCDs. In our quasi-backscattering configuration, the Poynting vector of incident light is along $[00\bar{1}]$ direction. Table below provides the selection rules in this configuration. The $E_{1g}$ or $E''$ modes can become active in the $\bar{Z}(X,Y)Z$ configuration due to a leakage of the polarisation along the [001] direction.

| Porto notation | 2H$_a$-NbSe$_2$ ($D_{6h}$) | 4H$_a$-NbSe$_2$ ($D_{3h}$) |
|:---:|:---:|:---:|
| $\bar{Z}(X,X)Z$ | $A_{1g} \oplus E_{2g}$ | $A'_1 \oplus E'$ |

| $\overline{Z}(X,Y)Z$ | $E_{2g}$ | $E'$ |

**Table 1** – The observable symmetries for a given scattering orientation of light with respect to the global symmetry of the lattices of 2H$_a$-NbSe$_2$ ($D_{6h}$) and 4H$_a$-NbSe$_2$ ($D_{3h}$).

### Heat capacity measurements

The AC heat capacity measurements in Košice in a magnetic field up to 8 T were performed using a chromel-constantan thermocouple. The thermocouple served as a thermometer and as a sample holder simultaneously, while the heat was supplied to the sample from LED via an optical fibre as described elsewhere[60]. Such configuration significantly reduces the addenda contribution to the total measured heat capacity and enables high-resolution measurement of the sample's heat capacity. However, the resulting data is in arbitrary units. The measurements were performed in a horizontal superconducting magnet allowing for controlled orientation of the crystal with respect to the magnetic field direction. The measurements in high magnetic fields up to 36 T were performed in Grenoble using a resistive chip. The sample was attached to the backside of a bare CERNOX resistive chip using a small amount of Apiezon grease. The resistive chip was split into heater and thermometer parts by artificially making a notch along the middle line of the chip. The heater part was used to generate a periodically modulated heating power $P_{ac}$. The induced oscillating temperature $T_{ac}$ of the sample was monitored by the thermometer part of the resistive chip. For details about the method see ref.[61]. The precise calibrations and corrections in magnetic fields were included in all measurements and the data treatment.

### First-principles calculations

The band structure calculation and fat-band analysis were performed using Quantum Espresso package[62]. The ONCV pseudopotentials[63] were used within PBE exchange-correlation functional[64]. The ground state calculation was performed using 21*21*1 k-points and electronic cutoff of 45 Ry. For the occupation of electronic states, we used Methfessel-Paxton[65] smearing with degauss value of 1 mRy.

### Acknowledgments

This work was supported by the projects APVV-23-0624, APVV-SK-FR-22-0006, COST action CA21144 (SUPERQUMAP), M.G. and T.M. acknowledge support by the Slovak Academy of Sciences project IMPULZ IM-2021-42 and project FLAG ERA JTC 2021 2DSOTECH. This work was supported by the Science Grant Agency project VEGA 2/0073/24. M.-A.M., Y. G and O. M. thank the European Research Council (ERC) under the European Union's Horizon 2020 research and innovation programme (Grant Agreement n° 865826).

### Author contributions

T.S., P.S. conceived the study, designed the scientific objectives and oversaw the experiments, S.S., L.C. synthesized the samples, D.V., B.S. carried out and analysed EDS experiments, D.V. carried out and analysed XPS experiments, J.B. performed XRD experiments, M.M, O.M., Y.G., D.V. performed and analysed Raman spectroscopy measurements, Z.P., J.K., C.M., T.K. carried out and analysed Heat capacity experiments, M.G., T.M. performed the first-principles calculations, T.S. wrote the manuscript with input from all co-authors.